# The peculiar light curve of the Symbiotic Star AX Per of the last 125 years

Elia M. Leibowitz and Liliana Formiggini

Wise Observatory and School of Physics and Astronomy, Raymond and Beverly Sackler

Faculty of Exact Sciences, Tel Aviv University, Tel Aviv 69978, Israel

elia@wise.tau.ac.il

# abstract

We analyze the last 125 years optical light curve of the symbiotic star AX Per through some remarkable correlations that we discovered in its power spectrum. The data were assembled from the literature and from the AAVSO database. A series of 6 major outbursts dominate the light curve. They are presented in the power spectrum as 13 harmonics of the fundamental frequency $f_a=1/P_a=1/23172$ day$^{-1}$. We refer to them as the "red" frequencies. Oscillations with the binary periodicity of the system $P_b=1/f_b=681.48$ d are also seen in the light curve, with particularly large amplitudes during outbursts. The $f_b$ peak in the power spectrum is accompanied by 13 other peaks on each side, to which we refer as the "blue" frequencies. A distinct structure in the frequency distribution of the blue peaks, as well as in their peak power are best interpreted as reflecting beating of the 13 "red" frequencies with the binary one. We suggest, following others, that the major outbursts of the system result from events of intense mass loss from the giant star. Mass accretion onto the hot component, partially through the L1 point of the system, took place in the last 125 years at a rate that oscillated with the 13 first harmonics of the $f_a$ frequency. The binary orbit is slightly eccentric and periastron passages induced modulation of the L1 accretion at the binary frequency. Hence the $f_b$ oscillations in the brightness of the star of amplitude that is modulated by the "red" frequencies of the system.

# 1. Introduction

AX Per is one of the well studied prototypes of the class of symbiotic stars. The stellar system consists of a giant of type M4.5 III (Murset & Schmid 1999) and a hot component, most probably a white dwarf, that are in a binary orbit around each other with a period of some 682 d,  This has been established photometrically as well as in radial velocity measurements (Mikolajewska & Kenyon 1992; Fekel et al. 2000).  A thick wind is emanating from the red giant. The gaseous medium is heated by the UV radiation of the hot component, producing the bright emission lines that characterize the spectrum of the object( Iijima 1988; Mikolajewska &Kenyon 1992; Ivison et al. 1993, Skopal et al. 2001).

The light-curve (LC) of the star is characterized by brightness variations of amplitude 0.1-0.4 mag alternating with epochs of activity characterized by variations reaching amplitudes of up to 3 mag. The binary period of the system has long been recognized as a major modulating agent of the optical output. The light oscillations are mostly attributed to the well known reflection effect in binary systems, by the giant star (Formiggini & Leibowitz 1990).  Skopal (1994) has shown that in the geometry of the system,  heating and ionization of the thick wind emanating from the giant by collision with the wind of the hot component and by its UV radiation are even more efficient in introducing variations with the binary cycle. In a few cycles of the binary period, eclipses of the hot component have also been identified (Skopal 1994, Skopal et al., 2001, 2011). These interpretations may be adequate for explaining the oscillations during quiescence states but not the wild fluctuations during the outbursts of the system. Also the energy source, as well as the physical process or processes that give rise to these outbursts are far from being understood.

More details about the present status of our knowledge about the system, observational as well as in theory, can be found in a number of excellent investigations in this system performed in the last 25 years (e.g. Mikolajewska & Kenyon 1992; Ivison et al. 1993; Skopal 1994; Skopal et al. 2001, 2011).

AX Per is one of the not too many stars the brightness of which started to be monitored systematically already in the 19th century. We are therefore fortunate in being able to construct the LC of the system that extends over the last 125 years. This work utilizes this rich treasure of data, so far not well exploited, in order to extract new knowledge about this system. This paper is the 7th one in our long lasting project of studying symbiotic stars through their historical light-curves (Formiggini & Leibowitz 1994, 2006, 2012 ; Leibowitz & Formiggini 2006, 2008, 2011).

## 2. Data

AX Per photometric behavior was first discussed by Lindsay (1932) who, analyzing Harvard plates, noticed variations with period P~ 650 d. The Harvard plates photographic magnitudes, digitized from the figure of Lindsay are our first set of data. The total time covered is from November 1887 up to June 1932, corresponding to JD 241 0590 - 242 6900. The data up to JD 242 3600 represent single observations, after this date they are 25 d means. From now on we omit the 2 first digits, 24, from the JD dates. We designate them as jd.

Another set of photographic data from the Harvard plates, based on a redetermination of the comparison stars magnitudes, covers the time interval from jd 23000 up to 31460 (Payne-Gaposchkin 1946). These are also presented as a plot, from which we extracted a second data set for our use.

Further photographic observations (Wenzel 1956) were retrieved from fig 4 of Skopal et al. (2000). This set is overlapped by a set of photographic estimates by the Association Francaise des Observateurs d'Etoiles Variables (AFOEV) from jd 32085 to 40290.

Eye-visual regular monitoring of AX Per brightness by the AFOEV and by the American Association of Variable Stars Observers (AAVSO) started around the year 1968. The data gathered by the AFOEV are consistent with those of the AAVSO, and in our analysis we used the AAVSO ones updated to May 5 2013 =jd 56417, binned into 30 d bins. We note that a few AFOEV data between jd 41272.5 and jd 41317.2 show a brightening of the star by more than one magnitude. Following an inquiry with the AFOEV, these points in the AFOEV LC should be discarded as spurious (E. Schweitzer, private communication). Jurdana-Sepic and Munari (2010) have published 6 B magnitude values and one V magnitude value extracted from historical plates of Asiago observatory between the years 1976 and 1985. Finally, we should note that the Mjalkovskij (1977) data were not used owing to the scale discrepancy shown both in the visual and in the photographic estimates.

## 2.1 Scaling the data

The photographic data, called from different papers and data bases, were checked for scale consistency. In fact, the new estimation of the comparison stars (Payne-Gaposchkin 1946) results in a difference of magnitudes between the two data sets that were retrieved from the Harvard plates. Analyzing the data after jd ~ 26000 when the system was in quiescence, we empirically estimated that the old Lyndsay's (1932) magnitudes are .24 mag fainter than those of Payne-Gaposchkin (1946). The Lyndsay's data were shifted and merged with the Payne-Gaposchkin's ones. As previously noted, the Wenzel (1956) data agree with the AFOEV photographic ones and the merging does not required any scale correction. The final data set of photographic estimates covers the time from jd=10590 to 40290.

The photographic set of data ends during a quiescence state of AX Per when also a few AAVSO data points are available. Using these short overlapping portions of the LC, the AAVSO data were shifted empirically, in order to connect the photographic and the eye-visual data. Fig. 1a presents the full LC of AX Per so obtained, covering the last 125 years of the star history. The y axis displays the magnitudes in normal units: nmag(t)=mean(m)-m(t), where m(t) are photographic magnitudes. The displayed LC is essentially the one presented in Skopal et al (2001, 2011), except for the omission of the spurious AFOEV points mentioned above, and the extension of the LC up to May 2013.

Figure 1b presents the outcome of an application of the running mean operator on the LC with a running window width of 682 days. We shall refer to this curve in Section 3.6. Frame c is the difference between the observed LC and the running mean curve. It shows in particular the strong correlation between the overall brightness of the system and the amplitude of its variation with the binary orbital frequency (see next section).

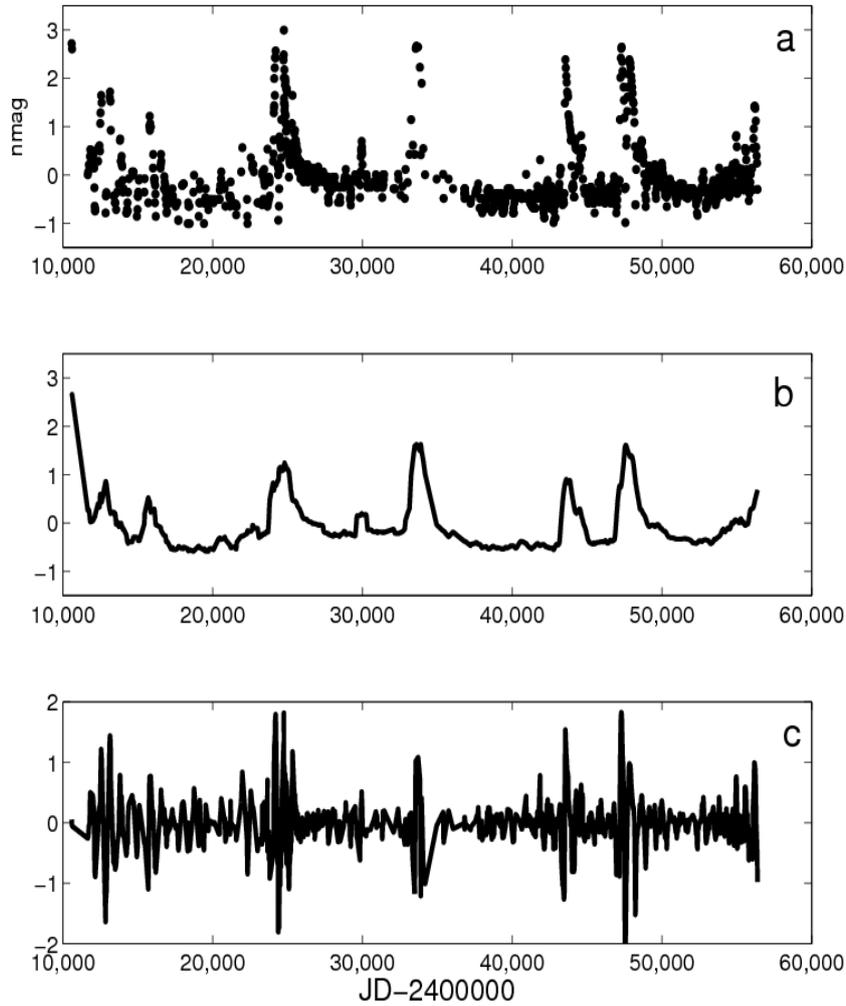

Figure 1: (a) Light curve of AX Per of the last 125 years. Magnitudes are in normalized units explained in the text. (b) Running mean of the light curve with running window of 682 day width. (c) Difference between the observed LC and the running mean one.

## 3. Time-series analysis

### 3.1 Power Spectrum

Figure 2a presents the power spectrum (PS) of the LC seen in Fig. 1a. Two clusters of peaks characterize the periodogram. One group is at the low frequency ("red") end, corresponding to periods of the order of 1700 d and longer. The second cluster ("blue") is centered around a distinct peak near the frequency $f_b = 1.47 \times 10^{-3}$ d$^{-1}$, corresponding to the well known binary period of the system. Here we adopt for this periodicity the value $P_b = 681.48$ d, a choice that will be explained in Section 3.5.

The left heavy vertical line marks the high frequency edge of a frequency band that contains the 13 high "red" peaks in the PS. Figure 2b is zoom on this passband in which the peak frequencies are marked by the short vertical lines. The two other heavy vertical lines in rame a mark respectively the low and the high frequency edges of two pass bands on the left and on the right of the binary frequency $f_b$, of width equal to that of the "red" band. Frame c is zoom on the frequency between these two vertical lines. The short low vertical lines mark the frequency $f_b$ and the peak frequencies of the highest 13 peaks in each of the two bands on the left and on the right of it. We refer to them as the b1 and the b2 peak sequences.

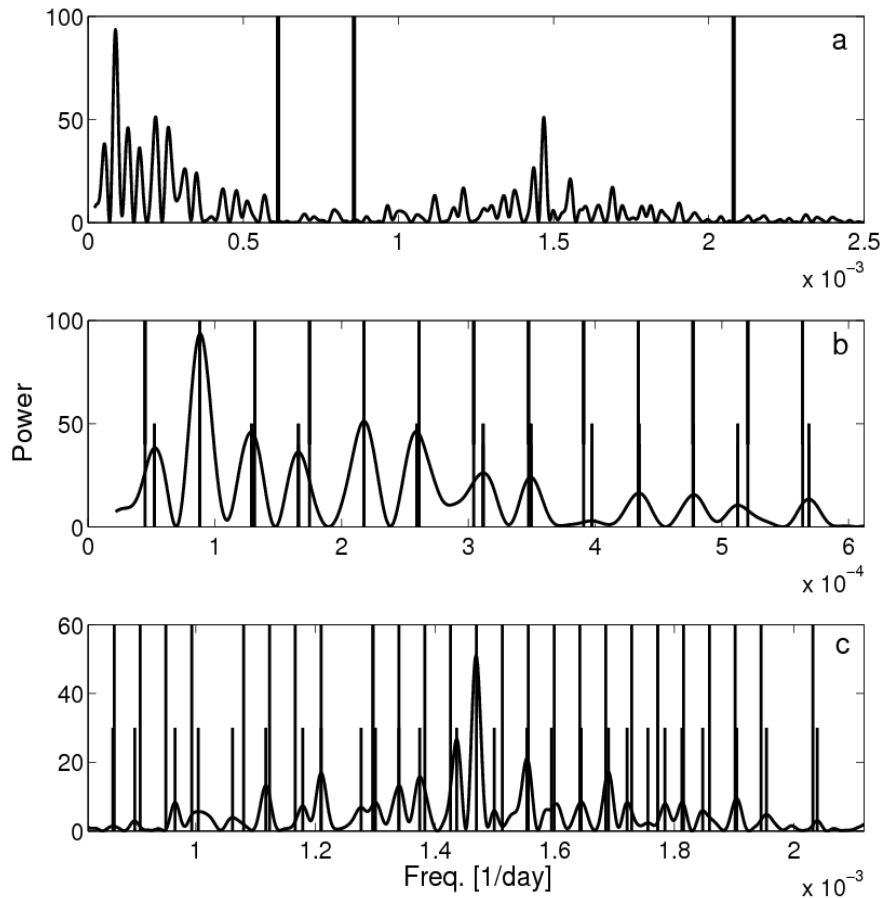

Figure 2: Power spectrum of the LC shown in fig. 1a. The meaning of the vertical lines is explained in the text.

## 3.2 The "red" cluster of peaks

Even an eye examination of Figure 2 reveals that to a very good approximation, the thirteen 'red' peaks are distributed along the frequency axis with a constant interval between any two neighbors. By the least squares procedure, with the peak power values as weights, we found the set of frequencies such that any 2 of them are separated by an integral number of a constant value, that best fit these 13 "red" peak frequencies. The constant value, termed "quantum" by Broadbent (1955) is found to be $f_a = 4.329 \times 10^{-5}$ $d^{-1}$, corresponding to the period $P_a = 23100$ d. The long vertical lines in the "red" band in Fig. 2b mark the frequencies $k/23100$ $d^{-1}$ for all integers $l <= k <= 13$. We adopt the value $P_a = 23172$ d for reasons explained in Section 3.5

In view of the above we shall refer to the 13 "red" peaks in the PS as harmonics of the fundamental frequency $f_a$. We note, however, that this does not necessarily imply that $1/f_a$ is a genuine periodicity of the AX Per system, since the observations cover no more than twice the length of this time interval.

## 3.3 The "blue" cluster of peaks

We applied the same least square procedure to find the "quantum", an integral number of which separates any two of the set of 27 peak frequencies in the "blue" PS of the star, marked by the vertical short lines in Fig. 2c. The "quantum" so found is again $f_a$, exactly the same as the one among the "red" sequence. The long vertical lines in frame c mark the frequencies of the best fitted set of frequencies with this "quantum" number.

We perform a bootstrap type test of the null hypothesis that the 13 peaks on each side of $f_b$ are distributed randomly along the "blue" section of the frequency axis of the PS. We consider the standard deviation S of the differences between the observed peak frequencies (short vertical lines in Fig. 2) and the predicted frequencies (long lines) as a measure of the quality of the least square fitting. Within each one of the two "blue" passbands around $f_b$ we created 16000 random distributions of 13 frequency values, with the constraint that no two numbers among them will be closer to each other than the theoretical resolution of the PS set by the total length of the LC. Applying the least squares procedure on each one of these distributions of 27 numbers we found only 249 cases with S value as small as, or smaller than the one associated with the 27 peaks in the PS of the observed LC. The probability that the observed 26 "blue" peaks around $f_b$ are due to random noise in the LC can therefore be rejected at a 98.4% statistically significant level. The equality of the "quantum" values in the "red" and in the "blue" sections of the PS reduces the probability of randomness in the distribution of the "blue" peaks to practically zero.

### 3.4 Correlation in power

We now consider the values of maximum power of the 26 peaks of the b1 and the b2 sequences defined above. We combine the powers of the nearest peaks to $f_b$ in the two sequences, then the powers of the second nearest peaks to $f_b$ in b1 and b2, etc. We thus obtain 13 values of power of the 13 pairs of peaks around $f_b$. Figure 3a is a plot of these 13 power values vs. the powers of the 13 "red" peaks seen in figure 2. The line in the figure is the regression line between these two sets of 13 power values. A bootstrap test on a sample 10000 sets indicates that the null hypothesis that the two sets are uncorrelated can be rejected at 99.96% probability. The slope of the line is 0.347 +- 0.168, where the quoted uncertainty indicates the 99% confidence interval around the slope value. Somewhat lower but still statistically significant correlations are found also between the 13 powers of the "red" sequence and the powers of the b1 and the b2 sequences considered separately.

Figure 3b demonstrates qualitatively the correlation that we find between the "red" and the "blue" sections of the PS. In it we plot again in dashed line the PS in the "red" passband. Superposed on it, in solid line, is a plot of the sum of the b1 and b2 sections, with $f_b$ as zero frequency, where b1 is considered in reverse order, namely from right to left. As in figure 2, the short vertical lines mark the peak frequencies of the 13 "red" peaks and the long ones mark the frequencies that best fit the "red" peaks with the $f_a$ quantum.

We emphasize at this point that these highly improbable features of the PS are uncovered in the periodogram of the raw LC of the star. In the process of their discovery we have made no assumption concerning the presence of any periodical content in the LC. Even the fact that $P_b$ is the well known binary period of the system has not been used in this analysis .

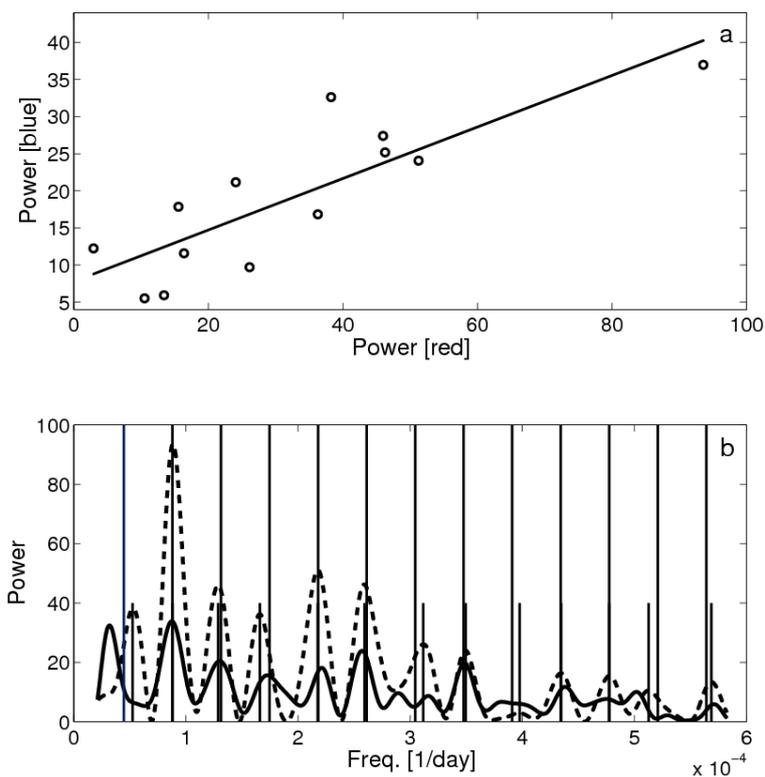

Figure 3: (a) Regression of the power of 13 pairs of peaks, one from the b1 and one from the b2 "blue" sequences peaks in the PS of AX Per, on the powers of the 13 "red" peaks in the PS.
(b) Dashed line - the "red" section of the PS of AX Per. Solid line - sum of the 2 "blue" sections on the two sides of the binary frequency in the PS, moved to origin of the frequency axis. The b1 section is considered in reverse direction. See text for further explanation.

## 3.5 Beats

The most natural and straightforward interpretation of the peculiar characteristics of the PS presented in the previous 2 sections is that they reflect beating of the 13 frequencies of the "red" sequence with $f_b$, the binary frequency of the system.

A light curve of oscillations at a frequency $f_b$ and at the 13 first harmonics of frequency $f_a$ that are also beating with the frequency $f_b$, can be expressed mathematically as follows

(1)
$$Y(t) = \sum_{j=1}^{13} A_j \sin[2\pi(jf_a t + \varphi_j)] + \sum_{k=1}^{13} B_k \sin[2\pi(kf_a t + \mu_k)] \sin[2\pi(f_b t + \theta_{mod})] + C\sin[2\pi(f_b t + \theta_{free})]$$

The first sum on the right hand side of the above formula represents free oscillations at the 13 first harmonics of the frequency $f_a$. The second sum represents the beats of these harmonics with the frequency $f_b$. We may also view this sum as representing an oscillator of frequency $f_b$, modulated by the 13 harmonics of $f_a$. The last term in the formula is required in order to produce in the PS the prominent peak of the $f_b$ frequency itself. This free $f_b$ oscillator is not necessarily in phase with the modulated one.

By least squares fitting of the Y(t) function to the observed LC we find the values of the two frequencies $f_a$ and $f_b$, the amplitude of the free oscillator and the phases of the two $f_b$ oscillators. The least squares procedure also yields the values of the amplitudes and phases of the 13 "red" harmonics, as well as those of the 13 harmonics of the same frequencies that modulate the modulated oscillator. Table 1 lists the values of $f_a$, $f_b$, the amplitude $A_{free}$ of the free oscillator, and the phases $\theta_{mod}$ and $\theta_{free}$ of the modulated and the free oscillators. Here phase zero is at jd 51135, the time of maximum radial velocity of the giant as reported by Fekel et al (2000).

|  | $P_a$ (day) | $P_b$ (day) | $A_{free}$ (mag) | $\theta_{mod}$ | $\theta_{free}$ |
|---|---|---|---|---|---|
| value | 23172 | 681.48 | 0.306 | 0.819 | 0.860 |
| +-uncertainty | 125 | 0.43 | 0.072 | 0.035 | 0.043 |

Table 1: Parameters of the function Y(t) that best fits the observed LC. They are the frequencies of the 2 dominant periods in the data, the amplitude of the free $f_b$ oscillator and the phase constants of the modulated and the free oscillators, relative to jd 51135.

The uncertainty values are estimated on the basis of bootstrap calculations. We consider the residuals of the observed minus theoretical Y value on each time point of the observations as a bank of "errors". We then create a pseudo-observed LC by considering the Y(t) values on each time point, to which we add one number chosen randomly from the error bank. We apply our least squares procedure on this pseudo-observed LC and obtain pseudo-observed values of the 5 parameters of Table 1. We perform these calculations on a sample of 200 such pseudo-observed LCs. The dispersion of the 200 values so obtained for each parameter defines the corresponding uncertainty number presented in Table 1.

The quality of the fit of the Y(t) function to the observed LC can be appreciated qualitatively by comparing the observed LC, displayed again as the upper curve in Figure 4, to the lower curve which is the best fitted Y function. The PS of the lower curve is almost indistinguishable from the PS shown in Figure 2.

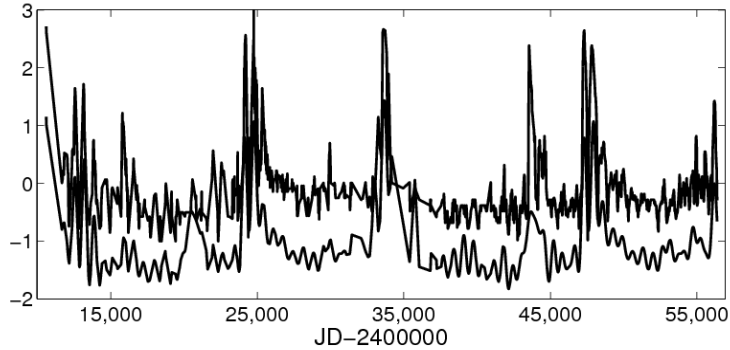

Figure 4: Upper curve - observed LC of AX Per. Lower curve - the Y(t) function, i.e. the analytical beats model, best fitted to the observed data.

In the least squares procedure, all the A and B parameters are treated as entirely independent, free parameters. We find that the resulting $B_k$ amplitudes are linearly correlated with the resulting $A_j$ amplitudes, as expected if the LC of the star does indeed represent physically beating frequencies.

We point out that our presentation of the theoretical LC in fig. 4 and the demonstration of its good fit to the observation are not intended to be regarded as evidence for the validity of our beating model. The figure serves only for showing that the LC of the star may well be simulated by a beating model with an appropriate choice of parameters. The evidence for the beats lies entirely in the peculiar features and correlations that we discover in the PS of the raw data that are clearly non random at a highly statistical significant level.

## 3.6 Comments on the fitted curves

1. The decomposition of the long term, outbursts structured variations of the LC into the 13 harmonic oscillations is of course implied by the PS but it is not an important part of the peculiarities of the PS. Nor is it essential in our discussion of them. We obtain practically the same theoretical LC and PS when considering a three component function consisting of the running mean LC shown in fig 1b, a free oscillator with the $f_b$ frequency, and a second oscillator of the same frequency modulated by the running mean, multiplied by a constant number. This is of course not an independent result but rather a consequence of the faithful representation of the running mean function by the 13 harmonics of $f_a$. In the least square fitting of this function to the data there are only 5 free adjustable parameters, namely the frequency $f_b$, the phases of the free and of the modulated $f_b$ oscillators, the amplitude of the free oscillator and the constant of proportionality mentioned above. The resulting $f_b$ value and the phases of the free and the modulated oscillators are practically identical to the numbers presented in Table 1. The amplitude of the free oscillator is also consistent, within the uncertainty interval, with the value given in Table 1. The constant of proportionality takes also a similar value to the slope of the regression line presented in Section 3.4.

2. The upper curve in Figure 5 is an extension of the PS of the observed LC beyond the frequency limit 1/450 $d^{-1}$ of the PS seen in figure 2, up to the limit 1/200 $d^{-1}$. This is roughly three times the mean Nyquist frequency of the LC time series. The lower curve is the same for the theoretical function Y(t). The theoretical function is of course entirely noiseless. It represents only the time variation that we have planted in it, namely, just the $f_a$ frequency with its 13 harmonics, the $f_b$ frequency and the corresponding beat frequencies. The features in the lower PS must therefore represent interferences among these frequencies, due to the final length of the LC and to the non equal sampling of it. This suggests that the nearly identical spectral features in the PS of the real data seen in the upper curve reflect similar interferences among the same basic frequencies, implying that these frequencies are indeed underlying the observed LC.

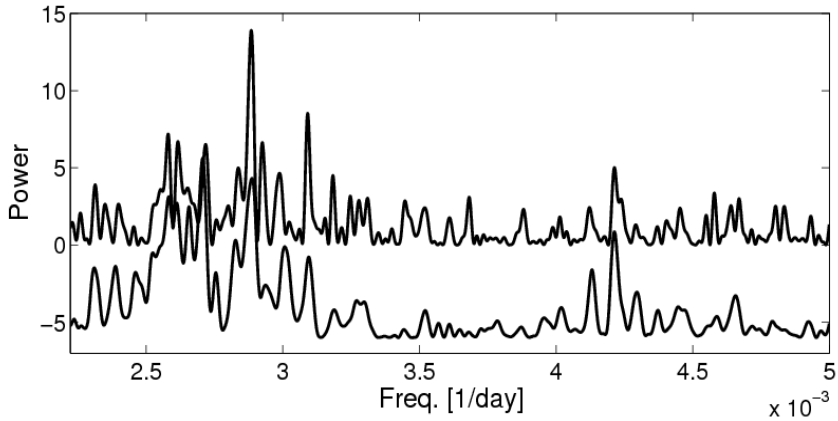

Figure 5: Upper curve -PS of the observed LC in the period range 450-200 days. Lower curve, the same for the best fitted theoretical Y(t) function.

3. Skopal at al (2012) performed precise photoelectric photometry of AX Per during some parts of the time interval jd $53000 < t < 56000$. They report on large oscillations of the system that are significantly out of phase of the $P_b$ periodicity that is prevalent almost always in the LC, at quiescence phases as well as during outbursts. These out of phase oscillations are also apparent in the AAVSO data. The getting out of phase of the $f_b$ variability during this time interval is well described by the theoretical Y(t), although not with the details of the structure that the LC assumed at that time. It is even predicted by the function Y(t) when fitted to the observed LC from which the data points of the above time interval are omitted. It seems that the out of phase variations at that time result from interference of frequencies involved with the Y(t) function.

## 3.7 Macro vs. micro presentation

Our present work is an attempt to understand the gross features of the optical photometric LC of AX Per. It addresses the time variations in the brightness of the star of the order of 0.5 mag and larger in brightness, and of time-scale of 450 d and longer. The system is varying also on shorter time-scale and with smaller amplitudes in brightness. The out of phase oscillations mentioned in the previous section is one example. Also Skopal et al. (2012) as well as the LC of the star presented in this work show that the $P_b$ periodicity is hardly being preserved on the "micro" scale. This is apparent in the varying shape of the $P_b$ cycle even at quiescence states of the star. It is also manifested by the fact that the time interval between apparent neighboring minima is varying by up to 100 days, and even more, around the $P_b$ periodicity. For example, the time intervals between the apparent minima of 7 consecutive $P_b$ cycles between jd 46500 and 51500 are 602, 678, 678, 693, 723 and 662 d. These different values are independent of the binning operation that is applied on the LC. We obtain nearly the same numbers when we consider the AAVSO LC binned into 10 or 30 d binning. Minima measured photoelectrically show similar excursions away from their expected times of occurrence. It should be noted, however, that some variation from a strict binary cycle may be attributed to variations in the ionization structure of the giant wind (Skopal 1998). In figure 1c one can also identify variability on time scale shorter and amplitude smaller than the variations on the binary cycle periodicity. The 681.5 d periodicity is, however, a robust feature of the system. The centers of the "micro" variability excursions preserve their periodicity and phasing throughout the last 125 years of the star history.

In this work we are concerned only with the 'macro' variability of the system. An underlying assumption in our approach is that the gross features of the LC carry information about the clocks that control the timing of the major events in the dynamics of the system. We assume that the effects of these clocks may indeed be decoupled from the micro-variations without losing this information.

All variations to which we refer as 'micro' have to do with the physical processes that transform the gravitational energy released in the accretion process and possibly also in nuclear reaction on the surface of the white dwarf to the optical output of the system. These are taking place in various components of the system such as winds of the two stars, an inter-binary and circum-binary nebulosity, gaseous streams and possibly also an accretion disk. They depend naturally on the dynamics and kinematics of these components, as well as on details of the thermodynamical processes within them. In view of their relatively smaller effects on the light curve these processes may be regarded as perturbations on the more orderly temporal behavior of the basic energy generation processes in the system. The detailed study of these perturbations is beyond the scope of this paper.

### 3.8 The relation with the spectroscopic ephemeris

If one assumes that the orbit of the AX Per binary system is circular, maximum radial velocity of the giant occurs a quarter of a cycle after the times of inferior conjunction of that star. In Section 3.5 Table 1 we reported that the phases of minimum light of the modulated and the free oscillators precede the phase of maximum radial velocity as given by Fekel et al. (2000) by less than 0.25. Note, however, that for $P_b$=681.48 d, the 2 very deep minimum points measured during two major outbursts, at jd 24389 and jd 47555 (see fig. 1a), are indeed at phase 0.75.

# 4. Discussion

In this section we propose a qualitative physical model for the AX Per symbiotic system that is likely to be a realization of the mathematical beating model suggested by the observations.

## 4.1 The Source of the continuum radiation

The main sources contributing to the optical continuum of AX Per are represented by the hot component, the giant and the nebula (Skopal, 2005). The hot component of the system and its close environment are undergoing partial and some time total eclipses by the giant of the system, giving rise to the modulations at the binary period of the system. These variations are also augmented by the reflection effect from the surface of the giant and from its thick wind (see Section 1). However, the periodic variations in the aspect ratio of the revolving AX binary system can hardly explain some basic observed features of the LC of the star.

As noted in Section 3.7, the observed times of minimum light of individual cycles make frequent excursions that may reach a distance of more than 100 days away from the nominal times of a strictly $P_b$ periodicity. In phase space these differences are sometimes as large as 0.15. It is hard to see how minima that are due to eclipses and to the reflection effect, being mostly geometrical in origin, can differ that much from the strictly periodic times of inferior conjunction. Furthermore, the amplitude of light variations due to an eclipse of the light source or to varying aspects of a reflecting medium of its light are to first approximation some given fraction of the source light intensity. The measured amplitude in magnitude units of variations due to these mechanisms alone should be roughly independent of light intensity of the source. This is clearly not the case in AX Per where the amplitude in magnitude units of the $P_b$ variations during outbursts may be 3 or more times the mean amplitude during quiescence (see figure 1c). Also the amplitude at quiescence intervals is also varying considerably with time.

The system must therefore include also an additional light source that is fluctuating at the binary frequency, nearly in phase with the eclipse and the reflection effects. The amplitude of this fluctuating source is strongly modulated by the same 13 "red" frequencies that gave the LC of the star the outbursts structure that it had in the last 125 years (Fig 1b).

## 4.2 Outbursts

The major energy source of the radiation of AX Per is in an accretion of material, from the wind and possibly also from the atmosphere of the giant star onto the hot component of the system. This process releases large quantities of gravitational energy. It may also supply fuel to an ongoing nuclear burning process on the surface of the white dwarf. . This has been proposed already more than 20 years ago by Mikolajewska & Kenyon (1992) and Ivison et al. (1993). The outburst episodes would then be events of extra material that is being ejected from the atmosphere of the giant star (Skopal et al 2011). The rate of mass loss from the giant is most likely controlled by an intrinsic parameter or parameters of this star. Skopal et al suggested, for example, that during one active phase of the system that they have studied, the radius of the giant increased by more than 10% (from 102 $R_O$ to 115 $R_O$). Variations in the mass loss rate may also be associated with variations in a global magnetic field in the outer layers of the star.

To date, there is no direct observational indication of a global magnetic field on the surface of the giant of AX Per.. However, non detection is by no means a detection of the absence (see for example the detection of magnetic field in Arcturus, the presence of which was unrecognized until very recently; Sennhauser & Berdyugina 2011). Cyclic or quasi-cyclic variability in the intensity, perhaps also in polarity, of stellar magnetic fields are ubiquitous, not only in our sun , but also in a wide range of different stars, including red giants (see for example Dal, Sipahi, Ozdarcan 2012, and many references therein). Also, in a recent paper, Olah et al. (2013) report on photometric observations in 3 K giant stars that vary on time-scale of years, possibly with periods of 5 or 10 years. These authors suggest that the photometric variability is a testimony on surface magnetic field variations of the same time-scale.

We note in this connection that in the series of papers of the present authors listed in Section 1, we have pointed out the possibility and produced at least circumstantial evidence for the presence of magnetic fields in the atmosphere of the cool components of a few well known symbiotic systems.

The outbursts structure of the LC is accordingly most probably an optical manifestation of variations in the value of the relevant intrinsic parameter of the giant star, which we assume for the rest of our discussion to be the radius of the star. The PS then shows that during the last 125 years the giant was pulsating as an harmonic oscillator of the 13 frequencies of the "red" peaks of the PS. According to Skopal et al (2001) if the inclination angle of the orbital plane is $i = 90^0$ the radius of a tidal lobe filling giant is $102\ R_O$ while if $i = 70^0$ the radius is $170\ R_O$. In fact, Mikolajewska & Kenyon (1992) have also suggested that the orbital inclination might be approx. 70 degrees with the red giant filling its tidal lobe. It would therefore seem that at each mode of the giant pulsations, accretion of matter through the L1 point of the system may have taken place. Alternatively or in addition, each mode of this pulsation created a corresponding mode of oscillations in the intensity of the wind emanating from the giant and consequently also in the rate of wind accretion of matter onto the hot component and also through the L1 point. The combined effect of the 13 frequencies oscillations in the accretion rate gave rise to the observed series of outbursts of the system.

## 4.3 The modulated oscillator

Accretion through the L1 point of the system is thus modulated with the frequency of any one of the star pulsation modes. If the binary orbit is non circular, the excess mass flow through the L1 point will be also modulated at the binary frequency. This will be translated to variations in the luminosity of the system at this frequency, with maximum light that is reached around a certain specific phase after periastron (Lajoie & Sills 2011). Details of the process of the release of energy, gravitational and/or nuclear, associated with varying accretion rate onto the hot component have been discussed by Sokoloski et al. (2006), in the context of another well studies symbiotic star, Z And. This mechanism may be suggested as the source of the very large amplitudes of the binary oscillations during outbursts.

We may describe the process in a slightly different language. Due to a slight eccentricity of the binary orbit, the mass accretion through the L1 point of the system is modulated at the binary frequency. The mass flowing towards the L1 point is supplied by the giant wind. Therefore the amplitude of $f_b$ oscillations in the accretion rate is modulated by the 13 frequencies of the giant pulsations. This is the physical origin of the second sum on the right hand side of equation (1).

Fekel et al. (2000) have analyzed all the radial velocity data for the giant star of AX Per that were available at the time of their work. They concluded that the binary orbit of this system is circular, following the precepts outlined by Lucy & Sweeney (1971). These authors have found however an orbital solution with a non zero eccentricity.

We have reanalyzed the radial velocity curve of AX Per as given in Table 8 in Fekel et al. (2000). Using formula (1) in Lucy & Sweeney (1971), we find a best fit orbital solution with eccentricity e=0.052. However, as noted by Fekel et al. we also find that due to the noise in the data, a circular solution is quite consistent with the data. In other words, the null hypothesis that the orbit is circular cannot be rejected by the presently available data. But as stated by Feigelson & Babu (2013), "... the null hypothesis can be rejected at a given level of significance, but the null hypothesis can not formally be accepted". Also in a recent paper Lucy (2013) has presented a new Bayesian inference for orbital eccentricities of single line spectroscopic binaries. One of his conclusions is that "Systems assigned e = 0 [according to these earlier tests] should preferably have upper limits $e_U$ computed". In Table A.3 in Lucy's paper it is also shown that upper limits for eccentricities could be significantly larger than the values derived by the least squares procedure. Thus a slightly non circular orbit of the AX Per binary system is certainly not ruled out by the presently available observations. Lajoie and Sills (2011) have shown theoretically that at least for some binary systems, an eccentricity of the order of 0.1 is enough for modulating an accretion process at the binary frequency.

## 4.4 The two oscillators

According to our proposition, two distinct oscillations are responsible for the variations of the brightness of AX Per at the binary period, the varying aspect ratio due to the system binary revolution and the varying mass flow through L1 due to the system eccentricity. This fact was so far elusive because the resulting two $f_b$ oscillations in the light of the system have nearly the same phase. The slight difference between these two phases, as presented in Table 1, could be real but analysis shows that it is statistically not significant.

The coincidence of near equality of the phases of the two oscillators may be regarded as a weakness of our model, but one has to bear in mind that such close phases occur with a probability that is no smaller than any other specific phase difference between the two. What enabled us to disentangle the effects of the two distinct oscillators in spite of their nearly equal phases is our analysis of the 125 years LC of the star in its entirety, and not only of a limited section of it, as was done so far by most studies of the system variability.

## 4.5 The free oscillator

The free oscillator term, the last one on the right hand side of expression (1), represents the variability that is due to eclipses and the reflection effect. The amount of the hot component light that is blocked by the giant, and the amount reflected from the giant surface or from any other fixed material element in the rotating binary system, at any given phase are determined primarily by the geometrical size of the giant star and perhaps also of an optically thick wind around it. To first approximation, these geometrical parameters are independent of the luminosity of the system. This is why this $P_b$ variability is represented by an un-modulated, free oscillator.

The phase of minimum light of the un-modulated oscillator that is due to eclipses should be the mark of inferior conjunction of the giant. Unfortunately, it cannot be determined by photometric measurements due to the effect of reflection that does not necessarily attain its maximum brightness level at superior conjunction of the giant. Also in general, the eclipse effect is being masked by interference with the oscillations of the modulated oscillator. The later is dominating the LC because of its larger amplitudes. The cycle of this oscillator is much less well defined on the "micro" level. This is due to the nature of the chain of physical processes that transform the variations in the gravitational/nuclear releases of energy to optical light fluctuations. Note, however, that when circumstances in the system allow identification of an eclipse, as is was probably the case at the two dates mentioned in Section 3.8, the eclipse had indeed preceded the phase of maximum radial velocity of the giant by quarter of a cycle.

## 5. Summary and conclusion

We discover a certain structure in the PS of the LC of the symbiotic star AX Per as well as a tight linear relation between the powers of its apparent peaks. These features are non random at a very high statistically significant confidence level. The outbursts of the system that characterize the long term structure of the LC can be decomposed into 13 interfering harmonic oscillations of periods of thousands of days. We suggest that these 13 "red" frequencies are those of oscillations in the value of some fundamental parameter of the giant atmosphere, a most likely one is the radius of the star. These oscillations drive oscillations at the same frequencies in the rate of mass loss from the giant and therefore also of the accretion rate onto the hot component. This is the origin of the series of the outbursts of the star. Some of the accretion onto the hot component is by overflow of material through the L1 point of the system. Due to a slight eccentricity of the binary orbit, this flow of matter is also modulated by the binary frequency. The beat of the 13 "red" frequencies with the binary one gives the "blue" section of the PS of the LC its observed peculiar structure. It also explains the linear relation that is found between the power of the "blue" peaks in the PS and that of the "red" peaks.

If the giant star continues to oscillate with the $f_a$ frequency also into the future, one could predict that the outburst that the system is presently undergoing will reach its peak luminosity sometime around jd 56840 (June/July 2014).

## Acknowledgments

We acknowledge with thanks the American Association of Variable Stars Observers for the data we extracted from their International Database that made this research possible. We also thank E. Schweitzer for providing us with useful information regarding the AFOEV data set. We also thank an anonymous referee for some useful suggestions.